\documentclass[acmtog,nonacm]{acmart}
\pdfoutput=1
\usepackage{booktabs} 
\usepackage{amsmath}
\usepackage{booktabs}
\usepackage{multicol}  
\usepackage{multirow}
\usepackage{subfigure}
\usepackage{amsmath}
\usepackage{array}

\usepackage[ruled]{algorithm2e} 

\SetAlFnt{\small}
\SetAlCapFnt{\small}
\SetAlCapNameFnt{\small}
\SetAlCapHSkip{0pt}

\acmJournal{TOG}

\setcopyright{none}

\begin{document}
\title{Binaural Audio Generation via Multi-task Learning}

\author{Sijia Li}
\affiliation{%
  \institution{Tianjin University}
  \city{Tianjin}
  \country{China}}
\email{lisj@tju.edu.cn}
\author{Shiguang Liu}
\authornote{Corresponding author, lsg@tju.edu.cn. }
\affiliation{%
  \institution{Tianjin University}
  \city{Tianjin}
  \country{China}
}
\email{lsg@tju.edu.cn}
\author{Dinesh Manocha}
\affiliation{%
 \institution{University of Maryland at College Park}
 \city{}
 \country{United States of America}}
\email{dmanocha@umd.edu}

\authorsaddresses{}

\begin{abstract}
We present a learning-based approach for generating binaural audio from mono audio using multi-task learning. Our formulation leverages additional information from two related tasks: the binaural audio generation task and the flipped audio classification task. Our learning model extracts spatialization features from the visual and audio input, 
predicts the left and right audio channels, and  judges whether the left and right channels are flipped.
First, we extract visual features using ResNet from the video frames. Next, we perform binaural audio generation and flipped audio classification using separate subnetworks based on visual features. Our learning method optimizes the overall loss based on the weighted sum of the losses of the two tasks. We train and evaluate our model on the FAIR-Play dataset and the YouTube-ASMR dataset. We perform quantitative and qualitative evaluations to demonstrate the benefits of our approach over prior techniques.
\end{abstract}

%
%
	


%
%


\begin{teaserfigure}
	\centering
	\includegraphics[width=7.1in]{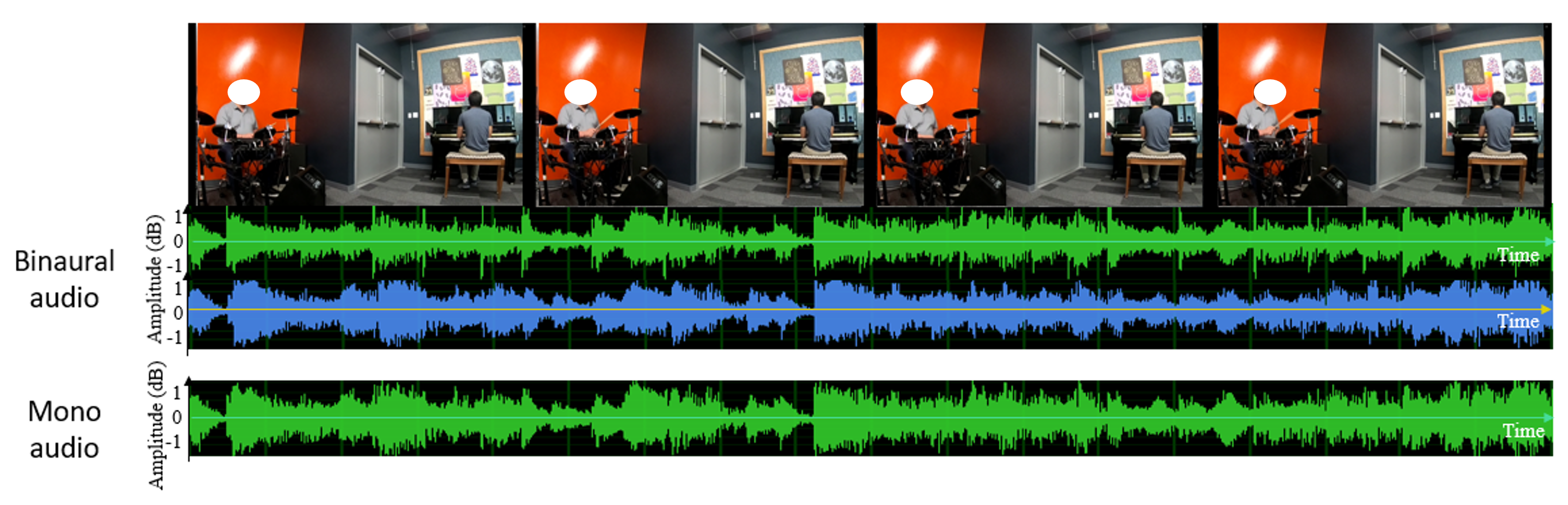}
	\caption{A video with synchronized binaural audio generated by our method. The top row highlights the video frames, and the bottom row highlights the mixed mono audio. The middle two rows are the corresponding left and right channels of the binaural audio. The two audio channels simulate the sounds for the left and right ears, thereby improving the sense of immersion.}
	\label{fig:head}
\end{teaserfigure}
\maketitle

\section{Introduction}

Humans interact with the surrounding world through a variety of senses. We can see objects in a scene, hear sounds, touch objects, etc. Specifically, vision and sound are two common modalities in our daily life. Audio-visual learning has gained more attention in recent years~\cite{zhu2020deep,morrone2019face,parekh2019identify,hao2018cmcgan,aytar2016soundnet,wiles2018x2face}, and many researchers have explored the relationships between visual and audio modalities. In particular, we can infer spatial information from both visual and audio data, which can help us locate objects in a scene and enhance our sense of scene understanding. Binaural audio, which is frequently used in virtual reality (VR) to enhance a user's sense of immersion, can imitate the sounds one hears in the physical world. As shown in Fig. \ref{fig:head}, the sound received at an individual's left ear is different from that received at the right ear. This topic has been well studied in spatial audio, and the resulting audio signals can be characterized by various parameters corresponding to inter-aural time difference (ITD) and inter-aural level difference (ILD), which are caused by the time difference and intensity of the sound received by our two ears \cite{rayleigh1875our,wightman1992dominant}. Based on the differences of the received audio signals between the two ears, we can judge the positions of sound sources.

Many videos recorded by users or available on the internet only consist of mono audio recordings. Moreover, these recordings do not carry enough spatial information to localize the sound source or provide immersive experiences. In order to capture accurate binaural sounds, we need two microphones attached to the two ears of a dummy head to record the audio. This may involve professional or expensive devices, which makes it difficult to capture accurate binaural audio data. In this work, we address the problem of recovering the spatial information from a given mono audio recording and reconstruct the corresponding binaural audio by leveraging the accompanying visual information. 

It is quite difficult to reconstruct spatial information with only a mono audio recording for all scenarios. To solve this problem, some works~\cite{morgado2018self,gao20192,kim2019immersive,zhou2020sep} estimate spatial information from visual information to aid the reconstruction of spatial audio. The audio and visual frames in the same video usually have a close relationship that can be exploited. 
  Morgado et al.~\shortcite{morgado2018self} use 360° videos as guidance to generate ambisonic sound. Kim et al.~\shortcite{kim2019immersive} reconstruct spatial audio by estimating room acoustics. \cite{yu2019self} add a classifier to generate stereo audio. Gao and Grauman~\shortcite{gao20192} present the FAIR-Play dataset, which contains thousands of normal-view videos with binaural audio recorded in a laboratory environment. 
  Zhou et al.~\shortcite{zhou2020sep} describe a new associative pyramid network for better audio-visual fusion. They treat the problem of separating two audio recordings as a special case of creating binaural audio and introduce additional videos of playing musical instrument scenes.

\begin{figure}[tb]
	\centering 
	\includegraphics[width=\columnwidth]{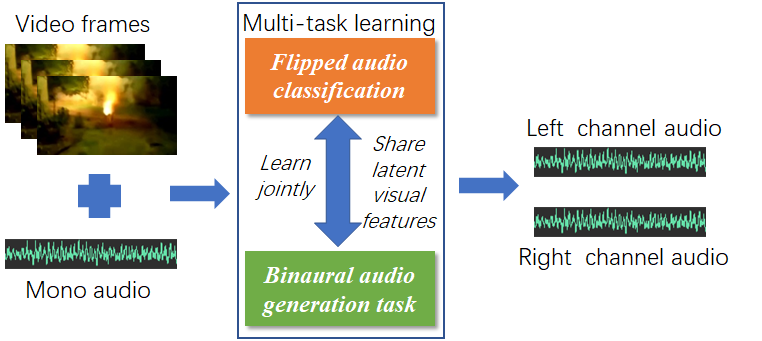}
	\caption{Problem formulation: The input to our learning algorithm is a mono audio recording and corresponding video frames. The output computed by our learning algorithm is binaural audio, including the left channel audio and the right channel audio. We adopt multi-task learning, i.e., through learning jointly and sharing latent visual representations between related tasks (the flipped audio classification task and the binaural audio generation task) to improve the performance.}
	\label{fig:sample}
\end{figure}



\noindent {\bf Main Results:}
 We present an approach based on multi-task learning to generate binaural audio from a video with mono audio recordings. Our approach is general, is applicable to all recorded videos, and does not need extra datasets. Instead of regarding separation of two audio recordings as a special case of binaural audio creation, we formulate it as a multi-task learning problem: the binaural audio generation task and the flipped audio classification task. By exploiting the ability of multi-task learning, we extract the latent information and share it between these two related tasks to improve the overall performance.
 
 
 We use ResNet \cite{he2016deep} to extract visual features from video frames and share them between the two tasks. Our subnetwork for binaural audio generation tasks aims to recover the spatial information from the mono audio based on extracting features from the visual frames.  We use the U-net architecture as the backbone network \cite{gao20192,zhou2020sep} and the associative pyramid network \cite{zhou2020sep} as a side-way network to generate binaural audio from mono audio. Moreover, our subnetwork for the flipped audio classification task is used to evaluate the consistency between visual and audio signals and whether the input left and right channels need to be swapped. Specifically, we use an encoder to extract representations of the randomly flipped audio and a classifier like \cite{yang2020telling} to distinguish it. These two tasks have some similarities, and we exploit the correspondence between the visual and audio modalities. We also account for the accuracy or sensitivity of the spatial information extracted from the visual frames. Therefore, we share the same visual network between them and add an attention network for each task to extract task-specific features from the shared representations.
Moreover, the flipped audio classification task can act as a regularizer for the generation task. This helps in terms of dealing with the issues of bias and overfitting and improves the performance. As a result, we do not need extra datasets and can handle general scenes (i.e., beyond musical instrument videos).
We train and test our model on the FAIR-Play \cite{zhou2020sep} and Youtube-ASMR \cite{yang2020telling} datasets. We evaluate our results both quantitatively and qualitatively.  We obtain scores of $0.846$ and $0.134$ for the STFT and envelope distance metrics, respectively. Our user study also shows that participants prefer our results (67\%) over results generated without multi-task learning (23\%) (stereophonic learning part in \cite{zhou2020sep}).
Our overall approach is highlighted in Fig. \ref{fig:sample}. The contributions of our work include:
\begin{enumerate}
\item We present an end-to-end approach using multi-task learning to generate binaural audio from a given video with a mono audio recording. We present an integrated method to solve both tasks corresponding to binaural audio generation and flipped audio classification.
\item Our learning solution for the classification task is designed to assist generation learning, which can check whether the visual information is consistent with the audio channels. Based on extracting sharing features and task-specific features between the two tasks, our model can learn more latent information from each task and thereby improve the generation performance. We achieve 0.846 and 0.134 for STFT and envelope distance on the FAIR-Play dataset, respectively (as compared to 0.879 and 0.135 in previous works.).

\item Our approach can generate comparable or better results without introducing additional datasets. We highlight the performance benefits on different datasets corresponding to playing musical instruments \cite{gao20192}, people talking, and rubbing objects \cite{yang2020telling}. In practice, our approach generates more realistic binaural audio without any additional data. In fact, we use less data. None of prior approaches combine the two tasks in this manner (Fig. 2). 

\end{enumerate}

\section{Related Works}
In this section, we give a brief overview of prior works on audio-visual learning, sound source separation, and audio spatialization.

\subsection{Audio-visual Learning}

Jointly learning the relationship between visual and audio modalities has become a popular topic in recent years~\cite{zhu2020deep,morrone2019face,parekh2019identify,hao2018cmcgan,aytar2016soundnet,wiles2018x2face}. By utilizing both audio and visual information, improved methods have been proposed for audio-visual generation~\cite{le2017generating,hao2018cmcgan,yalta2019weakly,wiles2018x2face}, audio-visual source separation and localization~\cite{Afouras2018,gao2018learning,senocak2018learning,tian2018audio,zhao2018sound,gao2019co}, audio-visual correspondence learning~\cite{hoover2017putting,nagrani2018learnable,afouras2018deep}, etc. Other methods focus on learning the general representation across visual and audio data with self-supervised learning and natural synchronicity ~\cite{aytar2016soundnet,hu2018deep} and spatial consistency~\cite{yang2020telling} between the two modalities. 
Techniques have also been proposed to generate a corresponding audio from the input video frames \cite{zhou2018visual,owens2016visually}. In contrast to these works, our goal is to convert mono audio into binaural audio, as opposed to directly generating binaural audio.


\subsection{Sound Source Separation}

Sound source separation, known as the “cocktail party problem” \cite{haykin2005cocktail,mcdermott2009cocktail,yost1997cocktail} has been extensively studied in signal processing and speech recognition. Prior works can be classified into audio-only methods and audio-visual methods.

\textbf{Audio-only methods.} Audio-only source separation is a challenging classic problem, and the goal is to separate the sound sources from a mixture. The major solutions are based on non-negative matrix factorization (NMF)~\cite{virtanen2007monaural,cichocki2009nonnegative}. Recently, many deep learning techniques have been proposed to solve this problem~\cite{wang2018supervised,chandna2017monoaural,yu2017permutation}. 

\textbf{Audio-visual methods.} Recently, audio-visual sound source separation techniques have been investigated~\cite{Afouras2018,gao2018learning,senocak2018learning,tian2018audio,zhao2018sound,gao2019co}. Compared with the audio-only methods, audio-visual methods use visual features to guide the separation, and these methods produce more cues. Zhao et al.~\shortcite{zhao2018sound} introduce the mix-and-separate training strategy and match the audio features with vision activations for each pixel. Gao et al.~\shortcite{gao2018learning} use traditional non-negative matrix factorization (NMF) methods to decompose the mixed audio and correlate it with the visual information. Gao et al.~\shortcite{gao2019co} detect visual objects and use the labels for supervision. Zhao et al.~\shortcite{zhao2019sound} incorporate the motion information into the network and suggest that motion cues can help to improve separation performance. Gan et al.~\shortcite{gan2020music} further add key features of the human body into the model. Our goal is similar to sound source separation. However, while sound source separation aims to separate sounds corresponding to different objects or sources, our goal is to separate sounds of the left and right channels.

\subsection{Reconstructing Spatial Information from Audio}
The user's immersion will be greatly enhanced if we can infer the location of the sound source from the audio signal. Many researchers solve this problem by using sound propagation techniques \cite{liu2020sound}. 
These techniques can be further classified into wave-based methods \cite{allen2015aerophones,mehra2013wave}, geometric methods \cite{krokstad1968calculating,cao2016interactive}, and hybrid methods \cite{yeh2013wave,rungta2018diffraction}. Many techniques have also been proposed to render spatial audio~\cite{schissler2016efficient,schissler2017efficient}. However, these methods need to first model the objects in the scene and separate sound sources to simulate the propagation effects. These methods are not directly applicable to recorded videos. 

Reconstructing spatial information from audio signals has recently been studied~\cite{morgado2018self,gao20192,kim2019immersive,zhou2020sep,li2018scene,lluis2021points2sound,xu2021visually}. Since mono audio lacks spatial information, it is difficult to reconstruct or extract the spatial information directly from a mono audio recording. Previous works have utilized the spatial information from the corresponding visual features or videos to aid the reconstruction. Recently, Morgado et al.~\shortcite{morgado2018self} proposed generating ambisonic audio from mono audio with the guidance of 360° videos. They predict the components of ambisonic audio and localize them in a visual scene. Kim et al.~\shortcite{kim2019immersive} propose a method to reconstruct spatial audio based on 360° videos by  estimating the room geometry and acoustic properties. Li et al.~\shortcite{li2018scene} also generate spatial audio using a recorded acoustic impulse response. Tang et al.~\shortcite{tang2020scene} present a deep learning method to estimate the acoustic material characteristics of rooms and use them for real-world audio rendering. All these methods have been used for real-world rooms and are not directly applicable to arbitrary video recordings. Lu et al.~\shortcite{yu2019self} use U-Nets to generate stereo audio via videos with normal fields of view. To further refine the model, they add a classifier to identify whether the left and right channels of the generated results are opposite. However, they add the classifier after the generation network and do not use multi-task learning. Yang et al.~\shortcite{yang2020telling} treat the binaural audio generation as a downstream task. They learn the audio-visual representation through judging whether the audio and video are spatially aligned. Then they use the pretrained features to improve the binaural audio generation. They consider the spatial alignment and audio spatialization separately and not as a whole. Gao and Grauman~\cite{gao20192} have developed a system to generate binaural audio from mono audio. They also present a dataset called the FAIR-Play dataset, which contains thousands of videos with binaural audio recorded under a laboratory environment. However, they use a simple scheme to fuse the audio and visual features. They did not explicitly associate the audio with the spatial information in video frames. Another work related to ours is \cite{zhou2020sep}. They regard the audio separation task as a special case of audio spatialization and develop an associative neural architecture to achieve better fusion. However, their approach requires extra solo music videos.



\begin{figure*}
	\centering
	\includegraphics[width=6.0in]{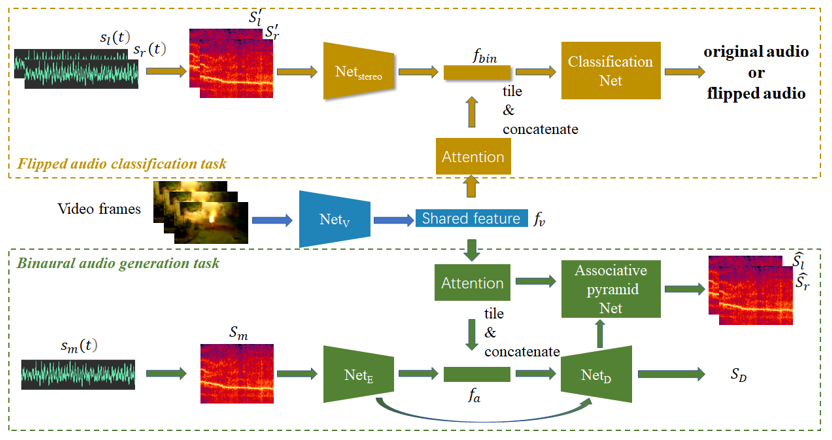}
	\caption{Overall architecture of our multi-task learning algorithm: We optimize two tasks at the same time: the binaural audio generation task (green) and the flipped audio classification task (yellow). They share the same visual network (blue part) and then extract task-specific features through the attention network. Therefore, latent visual information can be shared between them. We train the subnetworks to generate binaural audio and determine flipped (swap the left and right channels of the original audio) audio at the same time. We flip the left and right audio channels of the original binaural audio with a given probability and treat them as inputs to the flipped audio classification task. We optimize all subnetworks concurrently. Note that $s_m$ represents the mono audio waveform; $s_l$ and $s_r$ represent the left and right channel audio; $S_m$, $S_l$, and $S_r$ are the corresponding spectrograms; $S_D$ is the difference spectrogram of the left and right channels.}
	\label{fig:video}
\end{figure*}


\section{Our Approach using Multi-Task Learning}
\label{sec:sim}

Our goal is to convert a mono audio to a binaural audio for given video frames. In this section, we describe our overall multi-task learning approach in detail. The pipeline of our learning architecture is illustrated in Fig. \ref{fig:video}. We first present an overview of our overall architecture in Section 3.1. Next, we present the details of each model architecture in Section 3.2. Finally, we discuss the learning objective in Section 3.3.

\subsection{Overview}
\label{sec:framework}
In this section, we formulate our problem and give an overview of our architecture.

\textbf{Problem formulation:} Mono audio recording usually does not contain spatial information, and we cannot judge the direction of the sound sources only from that input. For our applications, our goal is to generate binaural audio containing sufficient spatial information and providing enough differences between the left and right channels. Our main goal is to recover the spatial information in the audio and can be represented as:
\begin{eqnarray}
Mono\rightarrow Binarual.
\end{eqnarray}
 It is impractical to generate realistic binaural audio directly from mono audio. Thus, we use the corresponding visual frames as guidance. Visual frames and corresponding audio describe characteristics of the same scene and contain complementary information. Therefore, we can utilize the spatial information in visual frames to perform audio spatialization: 
\begin{eqnarray}
Visual, Mono \rightarrow Binaural.
\end{eqnarray}
Our goal is to design a model to perform these computations.

Specifically, a binaural audio clip has two audio channels, that correspond to the sounds heard by our two ears, the left sound $s_l(t)$ and the right sound $s_r(t)$. The binaural audio provides supervision for the training process. We obtain the input mono audio by averaging the audio of the left channel and the right channel $s_m(t) = (s_l(t) + s_r(t))/2$. We also use visual frames $v$ as supplementary information to guide our algorithm. Therefore, we can formulate the problem as
\begin{eqnarray}
F(v,s_m)\rightarrow s_l, s_r,
\end{eqnarray}
where $F(.)$ represents the learning model.

\textbf{Multi-task Learning:} We solve this problem via \emph{multi-task learning}, which has been widely used in a variety of learning tasks~\cite{vandenhende2021multi,ruder2017overview}, including natural language processing~\cite{wang2018glue}, speech recognition~\cite{deng2013new}, computer vision~\cite{girshick2015fast}, etc. Instead of learning separate tasks through isolated networks, multi-task learning optimizes multiple tasks simultaneously. In this way, latent representations can be shared between related tasks, which can therefore improve the generalization and performance on the original task. Consequently, we formulate binaural audio reconstruction as a multi-task learning problem to improve the generation and performance.


\textbf{Architecture overview.} We design a model to solve this problem. As illustrated in Fig. \ref{fig:video}, our model mainly consists of two tasks: the binaural audio generation task (green) and the flipped audio classification task (yellow). For binaural audio generation, we utilize the U-Net structure (encoder $Net_E$, decoder $Net_D$, and skip connections) as the backbone network and the associative pyramid network to generate binaural audio. For flipped audio classification, we exploit the discriminator network ($Net_{stereo}$ and Classification Net in Fig. \ref{fig:video}) to determine whether the input audio channels are flipped. By training the two tasks at the same time, we can extract some latent information and share it between two related tasks. Overall, this approach works better than training with a single task. Moreover, we can learn a better representation for binaural audio generation because the other task, flipped audio classification, can act as a regularizer and reduce the risk of overfitting. 


Specifically, we extract visual features $f_v$ through a visual network (blue part in Fig. \ref{fig:video}), using a ResNet as our architecture to describe the video information. Then, for each task, we add an attention network as a feature selector to choose task-specific visual features. For all audio processes in our method, we operate in the frequency domain and transform the audio using Short-Time Fourier Transformation (STFT). We denote the frequency-domain spectrograms of the left channel and the right channel as $S_l$ and $S_r$, respectively. 
When loading the audio data, we randomly flip the left and right audio channels with a probability of $0.5$, and we set an indicator to indicate whether or not the audio is flipped.

The inputs of our model correspond to mixed mono audio, corresponding visual frames, and the randomly flipped binaural audio. We can finally recover the spatial information from the sound and estimate predicted binaural audio in a self-supervised way without introducing additional labels. Our model can be formulated as follows:
\begin{eqnarray}
f(v,S_m,S_l^{\prime},S_r^{\prime})\rightarrow \hat{S_l},\hat{S_r},y,
\end{eqnarray}
where $f(.)$ represents our networks, $v$ denotes the input video frames, and $y$ is the indicator. Note that $y$ equals 0 if the audio channels have been flipped, otherwise $y$ equals 1. $S_m$ is the mixed mono audio. We use $S_l^{\prime}$ and $S_r^{\prime}$ to represent the randomly flipped binaural audio. We also use $\hat{S_l}$ and $\hat{S_r}$ to represent the predicted binaural audio.

\subsection{Network Architecture}
In this section, we describe the main architecture of our approach, i.e., the visual network and the attention network, the binaural audio generation task, and the flipped audio classification task.

\subsubsection{Visual network and attention network}
Due to the lack of spatial information in mono audio, we need to learn this information from the corresponding video frames. Therefore, we utilize the visual network and the attention network to achieve this goal. To extract visual features, we use a modified ResNet18~\cite{he2016deep} as the architecture of our visual network $Net_v$ and pre-train it on ImageNet. We use the extracted visual features to assist the binaural audio generation and flipped audio classification learning process. Since our system needs to train two tasks at the same time, we add an attention network for each task after the visual network to extract task-specific visual features, which proved effective in \cite{Liu_2019_CVPR}. In other words, we use the visual network to extract the shared visual features. Next, we use the attention network as a feature selector to extract features related to the specific task based on the shared features.

We also perform ablation experiments to show the effectiveness of adding the attention network in multi-task learning, which will be discussed in Section 4.3. Specifically, for the attention network, given the visual features $f_v$, we add two convolutional layers and use a sigmoid to obtain a task attention map $a$. Finally, we multiply the shared visual features $f_v$ and the task-specific attention map $a$ to get the task-specific visual features. We use $f_{vb}$ and $f_{vf}$ to denote the specific visual features for the binaural audio generation task and the flipped audio classification task.



\subsubsection{Multi-Task Scheme}
In this section, we introduce the two tasks used in our method, the binaural audio generation task and the flipped audio classification task. As mentioned in Section 3.1, we can improve the performance through learning related tasks jointly. We choose these tasks because they have several similarities and also contain complementary information. Both of them have high requirements in terms of the spatial information and high correspondence between visual and audio modalities. Though they require different network structures due to differences with respect to the underlying tasks (regression task and classification task), they can share the same early processing layers (i.e., the visual network to extract visual features). Therefore, we can improve the performance by training the two tasks together. The latent complementary information is shared between them. 

\textbf{Binaural audio generation task.}
We generate corresponding binaural audio from mono audio through a backbone network and an associative pyramid network (APNet). 

\textit{The backbone network} is designed to convert mono audio to binaural audio with the guidance of visual information. For the audio spatialization, we use the conditional U-Net as the architecture, which is commonly used for audio processing \cite{gao20192,zhou2020sep,yu2019self}. The backbone network consists of two parts: the audio encoder network $Net_E$ and the audio decoder network $Net_D$. Further, $Net_E$ is designed to encode the input audio, and $Net_D$ is devised to reconstruct the binaural audio. We concatenate the visual features $f_{vb}$ at the bottleneck.

\textit{The associative pyramid network (APNet)} is developed to achieve a better fusion between visual features and audio features, as proposed in \cite{zhou2020sep}. It is a side-way network alongside the backbone network to implement the audio-visual fusion procedure. By associating different vision activations with audio feature maps, we can jointly learn audio-visual representations in a better manner. Thus, the visual information can guide the binaural audio generation more effectively.

\textbf{Flipped audio classification task.}
In addition to the binaural audio generation task, we perform this classification task. Our goal is to determine whether the input audio channels are flipped and thereby improve the performance of audio spatialization. Specifically, we randomly flip the left and right channels of the input audio. Judging whether the input audio channels are flipped means that we want our model to distinguish the original audio and the fake audio (flipped audio), in order to to learn the correspondence between visual and audio. This task is easier to learn and is related to the binaural audio generation task. To solve this task, we need to extract the visual features related to the left and right audio channels from the frames, which is also needed for audio spatialization. Both channels need to learn the spatial information in the video frame and the correspondence between visual and audio. This makes it possible to improve the performance by sharing latent complementary information and learning some features through one of the tasks. Moreover, adding this task can act as a regularizer and avoids overfitting. Specifically, we solve this classification task through the discriminator subnetwork. This task consists of two parts: the binaural representation network and the classification network.

\textit{Binaural representation network:} This network is designed to extract features to represent the binaural audio. As mentioned in Section \ref{sec:framework}, we randomly flip the left and right audio channels when loading the audio data. We stack the two audio channels and learn the overall binaural representation. The architecture of our binaural representation network is the same as $Net_E$ except that the input channel is double (two audio channels instead of one). The input of the network is the randomly flipped binaural audio. Through a series of convolution layers, the stacked audio channels are encoded into the binaural feature $f_{bin}$.  If we flip the input audio channels, we will get different binaural features. Therefore, we use this information to distinguish whether the visual information and audio channels are consistent.

\textit{Classification network:} Visual information and audio information should be consistent in the same video. This means that, if the sound source is on the left side of the screen, one should hear clearer sound in the left audio channel than in the right channel. For example, when a car moves from the left side of the screen to the right, one can also hear the sound moving from the left audio channel to the right channel. Therefore, it can be determined that the spatial information in vision data will be opposite to that in audio when the audio channels are flipped. The classification network is created to distinguish the flipped data from the original data. It contains a convolutional layer followed by a pooling layer. Then we add a fully connected layer and use a sigmoid as the activation. The input of the network consists of the visual feature $f_{vf}$ and the binaural feature $f_{bin}$. We use the visual feature as the guidance and concatenate it with the binaural feature as the final input audio-visual feature. Our network outputs an indicator (0 for flipped audio and 1 for the original) to show the classification results. Finally, we can use the output of the network to determine whether the audio channels are flipped.


\subsection{Learning Objective}

The final loss of our model is a combination of the losses of the two tasks. We calculate the losses of the binaural audio generation task and the flipped audio classification task. Then we obtain the weighted sum of these losses and optimize them as a whole. The selection and impact of different weights will be discussed in Section 4.3.3.

For the audio generation task, it is difficult to predict the final spectrogram directly. Instead of generating the spectrogram, we predict the mask as the common operation in sound source separation tasks. We evaluate the predicted spectrogram by multiplying the input spectrogram and the predicted mask. Therefore, for the binaural audio generation task, we follow the learning objectives in \cite{zhou2020sep} and predict the complex masks $M$=($M_{real},M_{imag}$), where $real$ and $imag$ denote the real part and the imaginary part, respectively. To handle complex spectrograms, compared with real spectrograms, we need to double the input and output channels of the audio network. Given a spectrogram $S_{input}$, the predicted spectrogram $S$ can be computed as follows:
\begin{eqnarray}
\begin{split}
S = S_{input} \times M = (S_{input}^{real} +j \times S_{input}^{imag}) \\
\times (M_{real}+j \times M_{imag}),
\end{split}
\end{eqnarray}
where $S_{input}^{real}$ and $S_{input}^{imag}$ denote the complex spectrograms of the input audio. All the spectrograms and masks used in our approach are complex numbers. For the flipped audio classification task, we adopt the cross entropy loss, which is usually used for classification. Therefore, the final learning objective of our approach consists of three terms: the difference loss $L_D$, the channel loss $L_C$, and the classification loss $L_{cls}$.

\textbf{Difference loss $L_D$.} The difference loss $L_D$ is the training objective of the backbone network. We calculate the difference spectrogram $S_D$ between the spectrograms of the left channel $S_l$ and the right channel $S_r$, which can be described as $S_D = (S_l - S_r)/2$. The input mono audio can be represented as $S_M = (S_l + S_r)/2$. This learning objective forces the network to learn the difference between the left and right channels. Therefore, the difference loss can be expressed as:
\begin{eqnarray}
L_D = ||S_D-S_D^{predict}||_2,
\end{eqnarray}
where $S_D$ is the ground-truth of the difference spectrogram and $S_D^{predict}$ is the prediction value.

\textbf{Channel loss $L_C$.} Instead of computing the difference spectrogram as the learning objective of the backbone network, we directly predict the spectrograms of the left channel $S_l$ and the right channel $S_r$ for the associative pyramid network. Thus, the channel loss $L_C$ can be expressed by:
\begin{eqnarray}
L_C = ||S_l-S_l^{predict}||_2 + ||S_r-S_r^{predict}||_2,
\end{eqnarray}
where $S_l$ and $S_r$ are the ground-truths of the spectrograms of the left and right channels, respectively; and $S_l^{predict}$ and $S_r^{predict}$ are the prediction values of the left and right channels, respectively.

\textbf{Classification loss $L_{cls}$.} For the classification task, we use the cross-entropy loss as our learning objective. We formulate the classification loss $L_{cls}$ as follows:
\begin{eqnarray}
L_{cls} =ylog \hat{y} +(1-y)log(1- \hat{y}),
\end{eqnarray}
where $y$ is the indicator value. Note that $y$ is 0 if the input audio channels are flipped, otherwise it is 1. $\hat{y}$ is the predicted value obtained by the discriminator network.

\textbf{Final loss $L$.} The final loss $L$ of our system is the combination of the difference loss $L_D$, the channel loss $L_C$, and the classification loss $L_{cls}$, written as:
\begin{eqnarray}
L = \lambda _1 L_D + \lambda _2 L_C + \lambda _3 L_{cls}.
\end{eqnarray}
where $\lambda _1$, $\lambda _2$, and $\lambda _3$ are the weights of the loss functions.\\[1em]

\section{Implementation and Performance}

In this section, we first introduce the datasets we use in our experiments in Section 4.1. Next, we discuss the implementation details for our system in Section 4.2. We highlight the quantitative evaluation results in Section 4.3 and qualitative evaluation results in Section 4.4.

\subsection{Datasets}
We evaluate our method on different datasets (as shown in Table \ref{tab:data}) and achieve state-of-the-art performance, which shows our approach works well for different types of audio samples. We train and test our model on the FAIR-Play dataset~\cite{gao20192}, which contains $1,871$ 10-second video clips recorded in a music room. All the videos are people playing instruments in a specific room with binaural audio attached. This is a clean dataset with high-quality binaural audio and little noise. We follow the splitting method proposed by \cite{gao20192}, i.e., 1,497/187/187 for training/validation/testing sets, respectively. 

We also test our method on a stereo dataset, YouTube-ASMR \cite{yang2020telling}, which is a large-scale dataset of ASMR collected from YouTube, with diverse spatial information. Therefore, this dataset is very suitable for evaluating our approach. It contains approximately $30,000$ 10-second video clips, and the total duration is about $96$ hours. We follow the dataset splitting provided by \cite{yang2020telling} (80\% for training, 10\% for validation, and 10\% for testing).

\begin{table} 
    \caption{Datasets and scenes used in our experiments}
    \centering
    \begin{tabular}{p{2cm}<{\centering} p{4cm}<{\centering}}
        \hline
        \hline
        \noalign{\smallskip}
        Datasets & Scene Characteristic   \\ 
        \noalign{\smallskip}
        \hline
        \noalign{\smallskip}
        FAIR-Play & Music  \\
       
        \noalign{\smallskip}
        \hline
        Youtube-ASMR       &  Human whispering  \\
        \noalign{\smallskip}
         & Human interacting with objects     \\   
        \noalign{\smallskip}
        \hline
        \hline
        \label{tab:data}
    \end{tabular}
\end{table}

\subsection{Implementation Details}
We sample the audio at $16$kHz and the video at $10$ fps. During the training procedure, we randomly sample audio with the length of 0.63s from 10s audio clips for training. We apply Adam Stochastic Optimization~\cite{kingma2014adam} with a learning rate of $0.0001$ for the visual network, the attention network, and the classification network in the discriminator subnetwork and $0.001$ for other parts of our system. The batch size is set to $16$. We follow the same configuration of the STFT computation (with window size 512 and hop length 160) and use the same size of sliding window for testing as \cite{zhou2020sep} during preprocessing and testing. All experiments were implemented using Pytorch. We trained and tested our model on an Nvidia RTX 2080. We trained $1000$ epochs for each experiment.\footnote{Our implementation is available at https://github.com/omeaningless/binaural-audio-generation.}

\subsection{Quantitative Evaluation}

%

\begin{table} 
	\caption{Quantitative comparisons between our method and previous methods on the FAIR-Play dataset. The results of \cite{gao20192} and \cite{zhou2020sep} are drawn from \cite{zhou2020sep}. For the results generated by our method, we calculate the average results across the 10 splits of the FAIR-Play dataset. We calculate the STFT distance and the envelope distance (ENV) as described in 4.3.1. We calculate the STFT distance and envelope distance (ENV) as described in 4.3.1.}
	\centering
	\begin{tabular}{p{4cm}<{\centering} p{1.5cm}<{\centering} p{1.5cm}<{\centering}}
		\hline
		\hline
		\noalign{\smallskip}
		Methods & STFT & ENV  \\ 
		\noalign{\smallskip}
		\hline
		\noalign{\smallskip}
		\cite{gao20192}        & 0.959  & 0.141  \\
		\noalign{\smallskip}
		\cite{zhou2020sep}  & 0.879 & 0.135  \\ 
		\hline
		\noalign{\smallskip}
		Ours & 0.846 &  0.134 \\ 
		\noalign{\smallskip}

		\hline
		\hline
		\label{tab:sigma}
	\end{tabular}
\end{table}


\begin{table} 
    \caption{Quantitative comparisons of ablation studies. Note that the backbone model has the same architecture as \cite{gao20192}. For more explanations for the models, please refer to Section 4.3.3.}
    \centering
    \begin{tabular}{p{5cm}<{\centering} p{1cm}<{\centering} p{1cm}<{\centering}}
        \hline
        \hline
        \noalign{\smallskip}
        Models & STFT & ENV  \\ 
        \noalign{\smallskip}
        \hline
        \noalign{\smallskip}
        Without multi-task learning & \\
        \hline
        \noalign{\smallskip}
        Backbone \cite{gao20192} ($L_D$) & 1.146 &  0.153 \\
        \noalign{\smallskip}
        APNet (stereophonic learning part in \cite{zhou2020sep}) ($L_D+L_C$)     &  0.919 & 0.139  \\
        \noalign{\smallskip}
            Training two tasks separately & 0.934 &  0.140 \\
        \noalign{\smallskip}
        Classifying the flipped generated results & 0.935 &  0.140 \\ 
        \noalign{\smallskip}
        \hline
        \hline
        With multi-task learning & \\
        \hline
        Without weighted loss and attention ($L_D+L_C+L_{cls}$) & 0.922 &  0.140 \\
        \noalign{\smallskip}
        Without attention & 0.869 &  0.136 \\
        \noalign{\smallskip}
        Our whole method with multi-task learning ($\lambda_1L_D+\lambda_2L_C+\lambda_3L_{cls}$) & \textbf{0.863} &  \textbf{0.135} \\
        \noalign{\smallskip}

        \hline
        \hline
        \label{tab:sigma}
    \end{tabular}
\end{table}

\begin{figure*}
	\subfigure[]{
		\label{fig:subfig:3a} 
		\includegraphics[width=6 in]{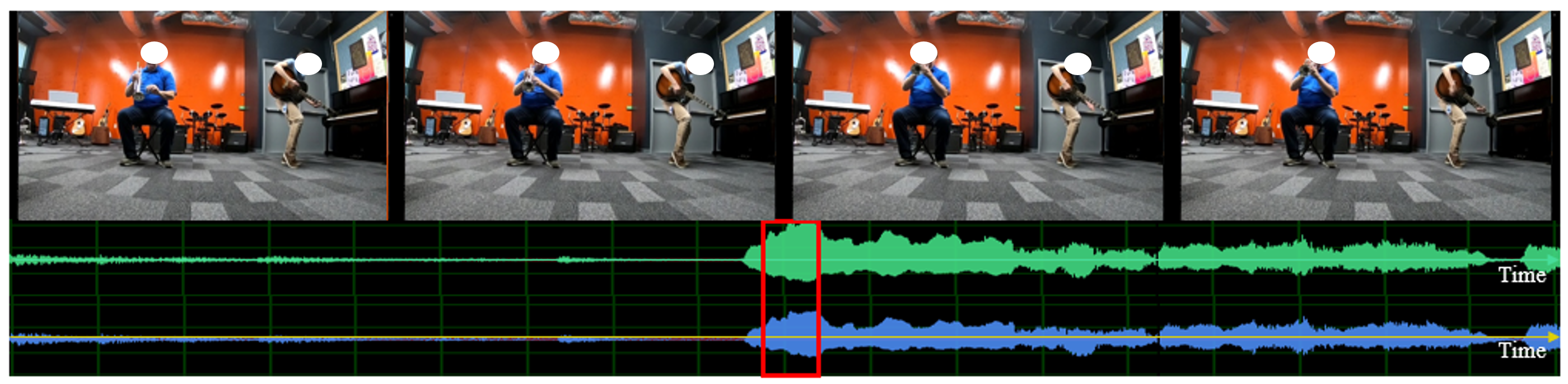}}
	\subfigure[]{
		\label{fig:subfig:3b} 
		\includegraphics[width=6 in]{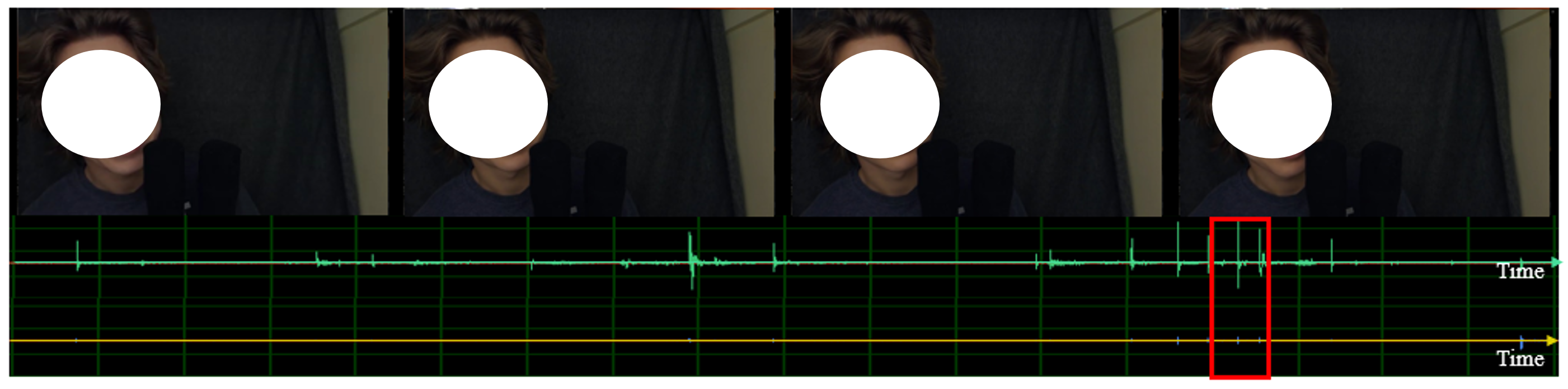}}
	\subfigure[]{
		\label{fig:subfig:3b} 
		\includegraphics[width=6 in]{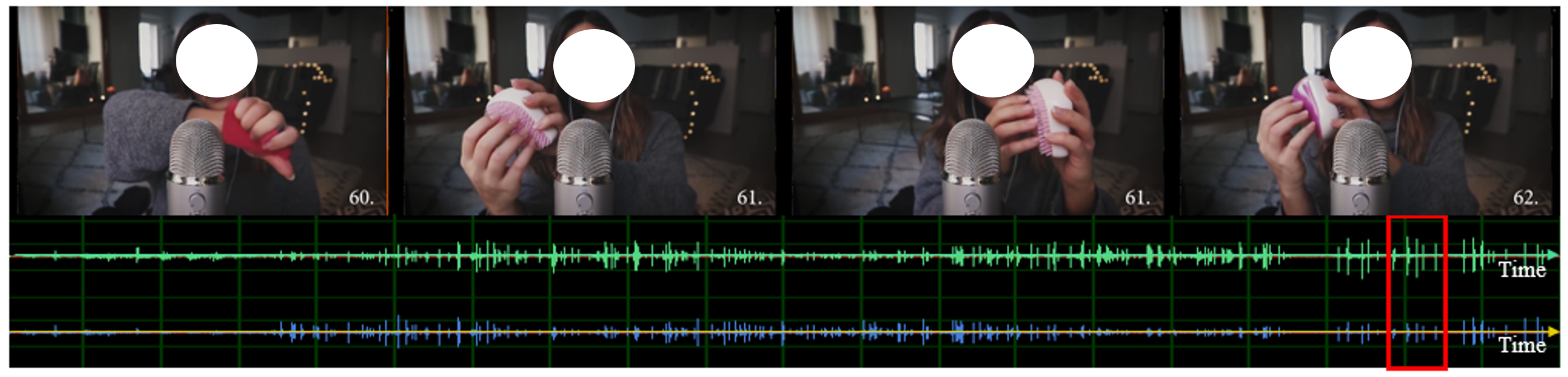}}
	\caption{Results on the FAIR-Play (a) and YouTube-ASMR ((b) and (c)) datasets. (a) is a music room scene. (b) is a scene with a person whispering. (c) is a scene of a person interacting with objects. We show the audio samples generated by our method. For each audio result, the top row is the left audio channel and the bottom row is the right audio channel. We can observe the differences between the two channels, as highlighted by the red boxes.}
	\label{fig:r2}
\end{figure*}
\begin{figure*}
	\subfigure[]{
		\label{fig:subfig:3a} 
		\includegraphics[width=6in]{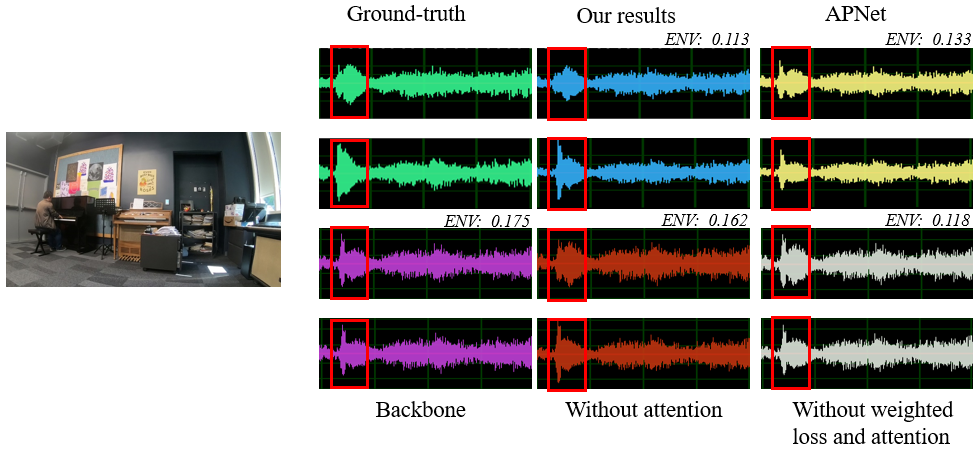}}
	\subfigure[]{
		\label{fig:subfig:3b} 
		\includegraphics[width=6in]{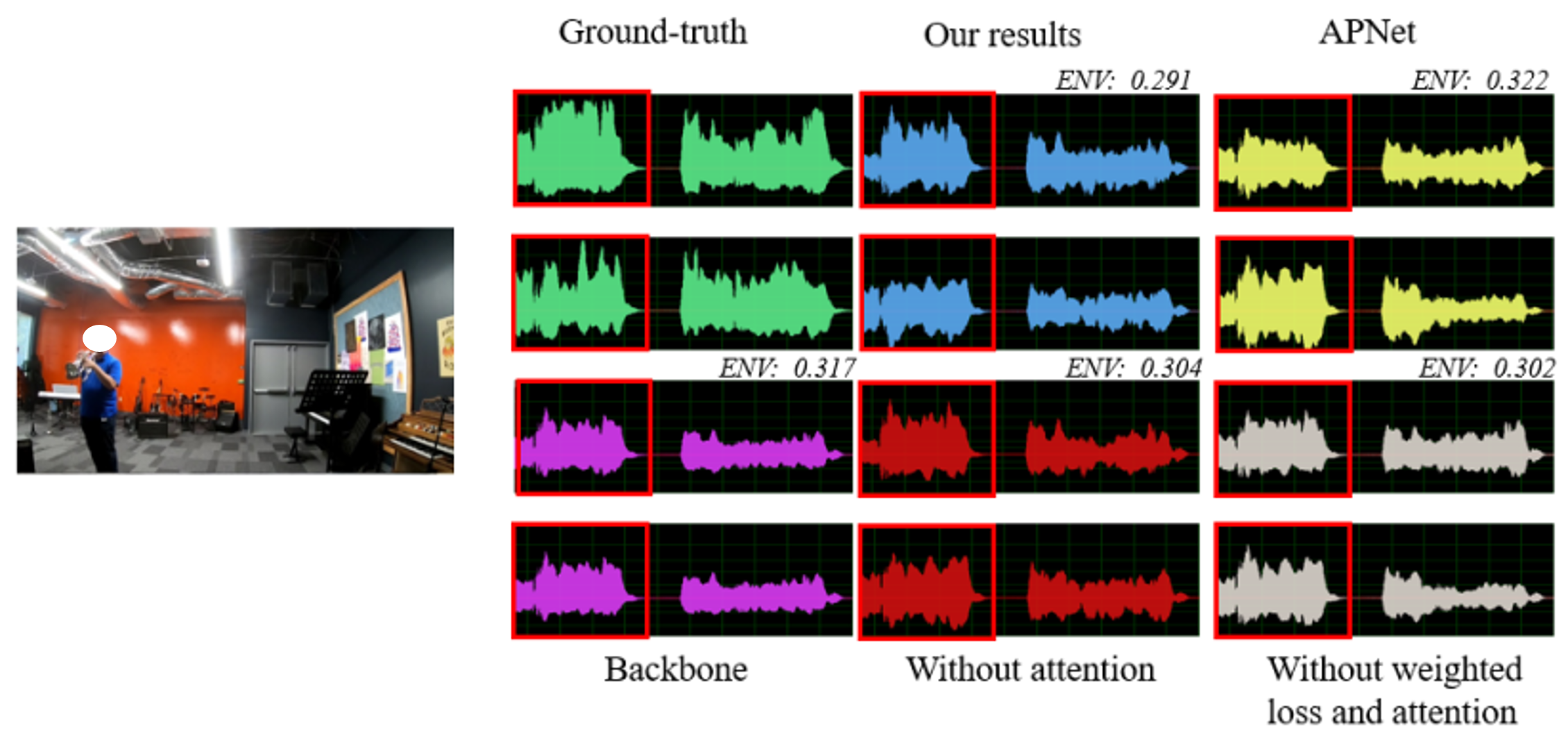}}
	\caption{Comparisons with different network architectures on the FAIR-Play dataset. For each audio result, the top row is the left audio channel and the bottom row is the right audio channel. We show the waveforms of the ground-truth (top left), audio waveforms generated by our method (top middle), APNet (top right), backbone (bottom left), without attention (bottom middle), and without weighted loss and attention (bottom right). In particular, the backbone network has the same architecture as \cite{gao20192}. Backbone and APNet are trained on the single task. Our results with multi-task learning are more similar to the ground-truth, both for the shape of the waveform and for the envelope distance. 
	The obvious differences are indicated in the red boxes. Note that we calculate the envelope distance (ENV) between the generated results and the ground-truth due to its ability to capture perceptual information.}
	\label{fig:r1}
\end{figure*}

\begin{figure*}
	
		\label{fig:subfig:3a} 
		\includegraphics[width=6in]{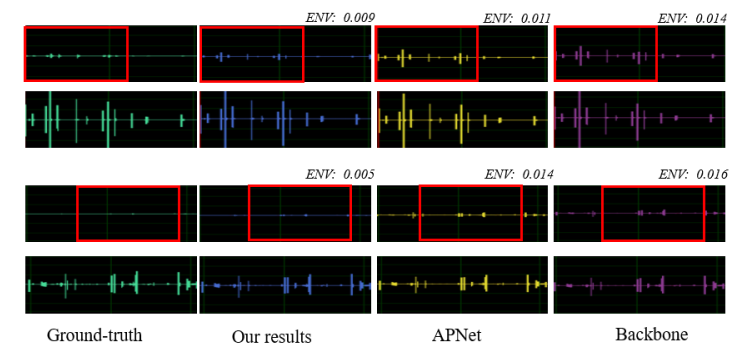}
	\caption{We highlight the performance with different network architectures on the ASMR dataset. For each audio output, the top row is the left audio channel and the bottom row is the right audio channel. We show the waveform corresponding to the ground truth, and compare with the waveforms generated by our method, APNet, and Backbone. Our results with multi-task learning are closer to the ground truth. }
	\label{fig:asmr}
\end{figure*}

\begin{figure}
	\subfigure[]{
		\label{fig:subfig:3a} 
		\includegraphics[width=2.2in]{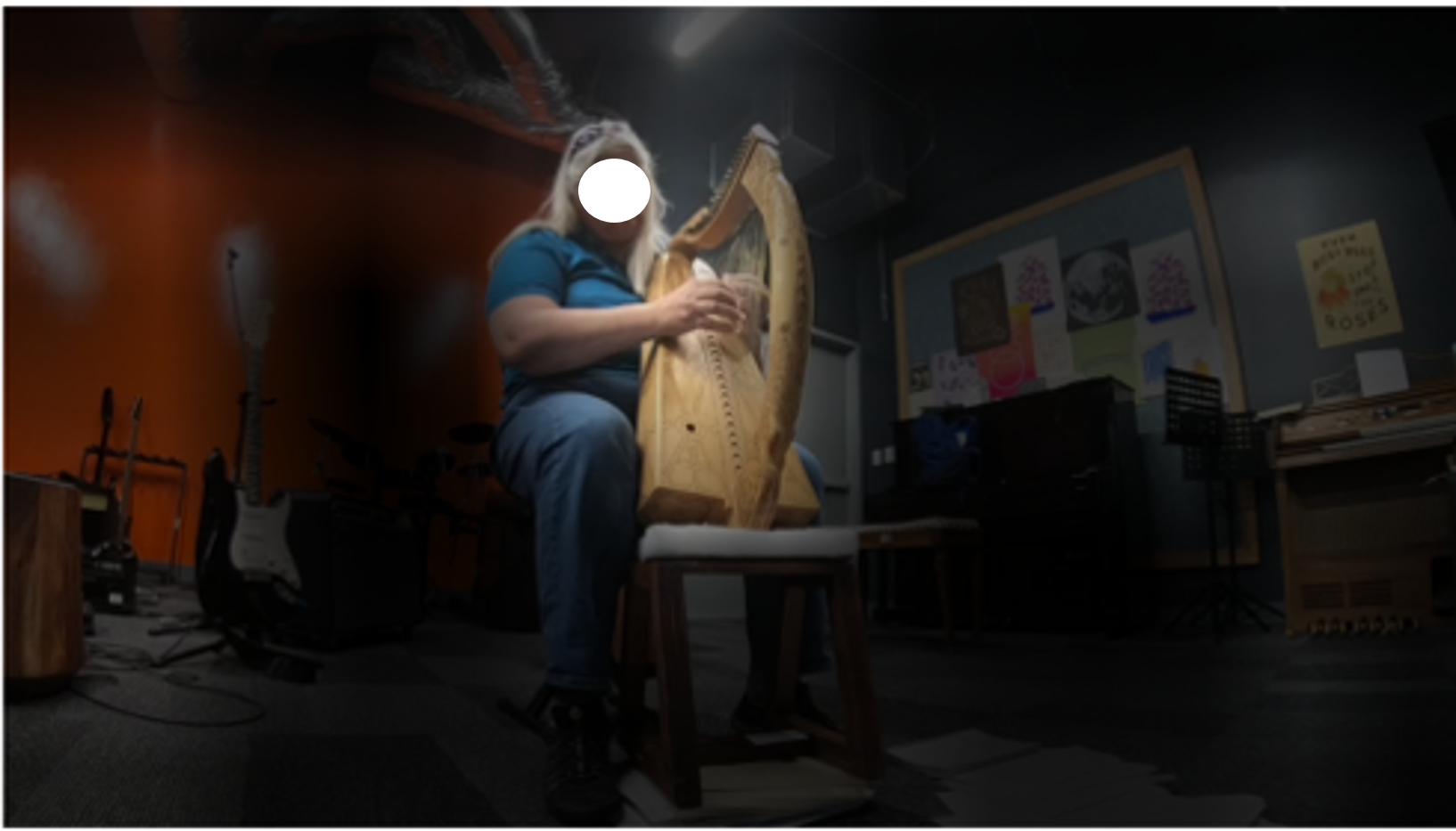}}
	\subfigure[]{
		\label{fig:subfig:3b} 
		\includegraphics[width=2.2in]{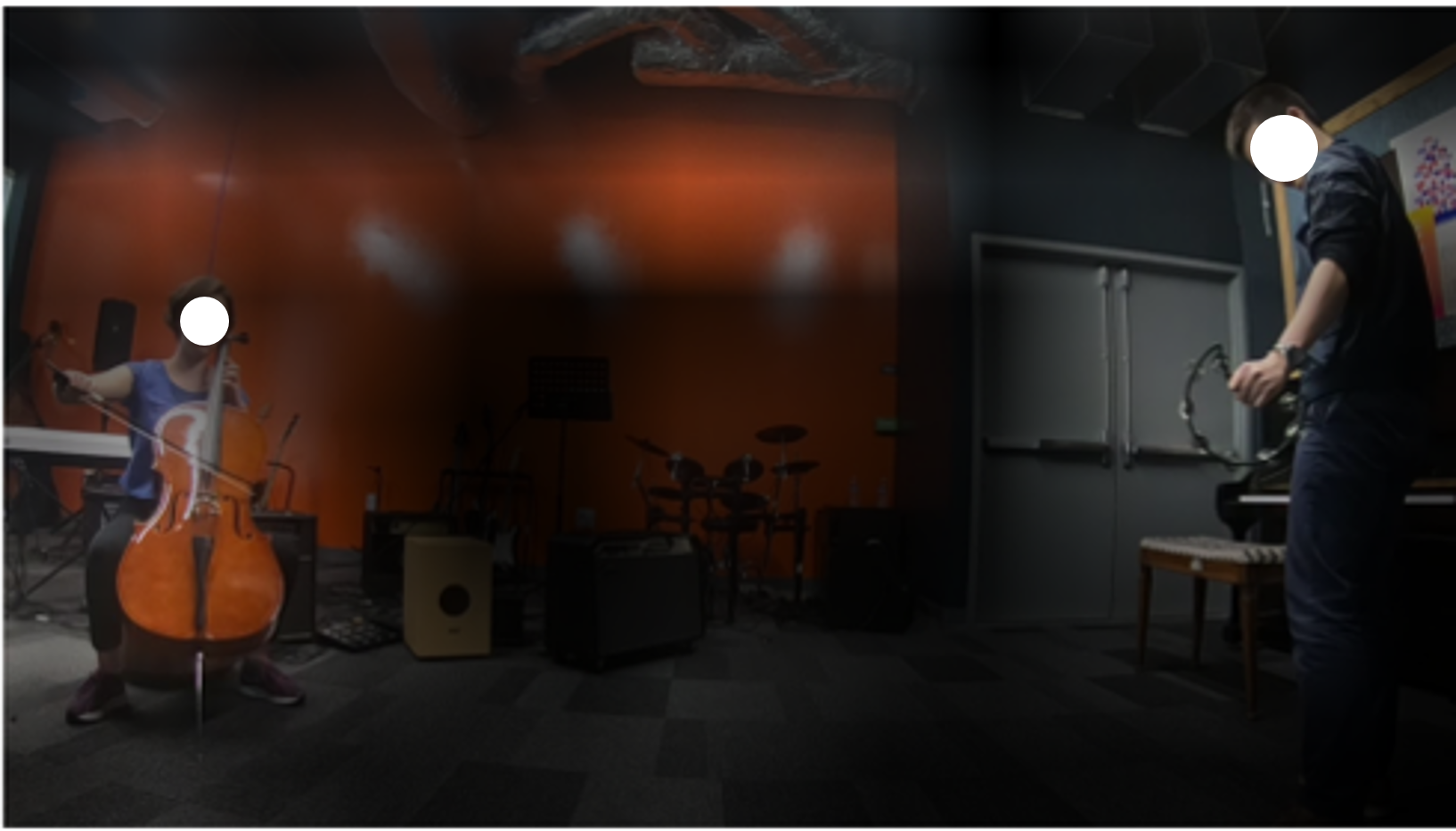}}
	
	\caption{Visualization of salient areas of the input frames. We normalize and resize the visual features in the binaural audio generation task $f_{vb}$ to the image size and overlay them on the original frames. This shows which part of the image is important for the audio generation procedure. (a) is a case of single source and (b) is two sources. The bright areas are the salient regions, and the network will focus more on these areas. We find that the areas of instruments are more salient, which means more attention will be paid to the instrument areas in the frames during binaural audio generation. This is consistent with our intuition. }
	\label{fig:vis}
\end{figure}

\subsubsection{Evaluation metrics}
We use well-known metrics in this field \cite{morgado2018self,gao20192,zhou2020sep} to evaluate the performance of our generated binaural audio. There are two kinds of metrics used in our experiments: the STFT Distance and the Envelope Distance (ENV). \\[1em]

\textbf{STFT Distance.} We compute the Euclidean distance between the predicted spectrograms and the ground-truth to illustrate the effectiveness of our results. This measure can be formulated as follows:
\begin{eqnarray}
D_{stft} = ||S_l-S_l^{predict}||_2 + ||S_r-S_r^{predict}||_2,
\end{eqnarray} 
where $S_l$ and $S_r$ separately correspond to the ground-truth of the spectrograms of the left channel and the right channel, and $S_l^{predict}$ and $S_r^{predict}$ are the predicted values (from a learning method) of the left and right channels, respectively.\\[1em]

\textbf{Envelope Distance.} We also calculate the Euclidean distance between the envelopes. Compared with the raw waveform, it is more suitable to use envelope distance as the index because it can capture the perceptual similarity in the audio samples~\cite{morgado2018self}. It can be described as 
\begin{eqnarray}
D_{env} = ||E_l-E_l^{predict}||_2 + ||E_r-E_r^{predict}||_2,
\end{eqnarray}
where $E$ represents the envelope of a signal. $E_l$ and $E_r$ are the ground-truths of the envelopes of the left channel and the right channel, respectively.  $E_l^{predict}$ and $E_r^{predict}$ are the predicted values based on a learning method.

\subsubsection{Comparisons}

In this section, we compare our results with the state-of-the-art methods~\cite{gao20192,zhou2020sep} and also describe our ablation studies. 
Comparisons with the previous methods~\cite{gao20192,zhou2020sep} are shown in Table 2. 
We train and evaluate the models on the FAIR-Play dataset and report the average values across the 10 splits provided by \cite{gao20192}. Note that \cite{gao20192} only utilize a U-Net backbone network, and \cite{zhou2020sep} also use the APNet. Moreover, \cite{zhou2020sep} performs sound source separation as an additional task, and this introduces additional mono data with a single sound source due to the mix-and-separate strategy \cite{zhao2018sound}. They use the solo part of the MUSIC dataset \cite{zhao2018sound}, which usually involves a single person playing an instrument, to perform the separation learning.  The results can be found in Table 2 \cite{zhou2020sep}. The qualitative results of the stereophonic learning part in \cite{zhou2020sep} under our experiment settings can be found in Fig. 5 and Fig. 6. 

We compare our method with \cite{gao20192}, which has the same architecture as our backbone network. 
These results contain some spatial cues, but still not enough. 
We also compare our method with \cite{zhou2020sep}. The use of APNet and extra data makes the performance better than \cite{gao20192}. 
The bottom row in Table 2 contains the results of our method. 
It can be observed that our method gets the best scores without using any additional training data.

\subsubsection{Ablation studies}
We also perform some ablation studies. The results are collected in Table 3. All experiments were configured as described in Section 4.2. We trained and evaluated all models on the splitting manner (i.e., split1) provided by \cite{gao20192} of the FAIR-Play dataset (it has 10 splits of the dataset in all). For each experiment, we train and test three times and then calculate the average results. Each model is trained for 1000 epochs.

First, without our multi-tasking scheme, we only use the backbone network, which is the same as \cite{gao20192}. We notice that results generated by this method get the worst scores. Next, we add APNet, which has the same architecture as the stereophonic part described in \cite{zhou2020sep}. This improves the results. Further, we add the flipped audio classification task and perform multi-task learning. We compare with the results generated by APNet, which is trained for a single task. Our overall method generates better results than prior approaches.


\noindent{\bf Weighted Losses:} We also evaluate the effect of choosing different combinations of the weights of losses in our multi-task learning approach. Instead of using the same weights for all losses (without weighted loss and attention in Table 3), we choose the best combination based on the gradient value of convergence for each task (without attention in Table 3). We observe that different losses in our tasks usually have different learning speeds of convergence. 
To balance between different losses, we train each task separately and observe the mean gradient values. 
We choose the appropriate coefficient as the weight of corresponding loss to make the two tasks have similar gradients at the end of training. Finally, we add the attention network and obtain the best scores. We set $\lambda_1$ and $\lambda_2$ as 44 and $\lambda_3$ as 1 in our final version, based on the experiment settings in Section 4.2.

\noindent{\bf Benefits of Multi-Task Learning:} 
We verify the effectiveness of the multi-task learning strategy in our model. We train the two tasks separately in order (shown as "training separately" in Table 3) and check whether this strategy performs worse than our method. Specifically, we train the flipped audio classification task first. Next we use the pre-trained visual network to initialize the visual network in our binaural audio generation task. We also perform another experiment in which we classify the generated results (also with two channels and can be flipped) instead of the input binaural audio (shown as "classify the generated results" in Table 3). That is to say, we move the discriminator network after the binaural audio generation task. The inputs of the discriminator network are the original, and the flipped results are generated by the binaural audio generation task. 
We classify the generated audio as "real." Next, we exchange the left and right channels of the generated audio and classify it as "fake." Finally, we observe from Table 3 that our whole method based on multi-task performs the best. The above comparisons indicate that we can achieve gains by our multi-task learning method. We have also compared the backbone network and APNet with our method on the Youtube-ASMR dataset. We achieve STFT distance 0.222 and envelope distance 0.058 for the backbone network, 0.206 and 0.056 for APNet, and 0.190 and 0.053 for our method, which also shows the effectiveness of our method.

\subsection{Qualitative Evaluation}
In this section, we present details of our qualitative evaluation.

\subsubsection{Qualitative results}
Although we use quantitative metrics to evaluate the generated audio, we still need to observe the results in a qualitative manner. Designing metrics consistent with human auditory perception is still a challenging problem. Therefore, we directly render our generated results to evaluate the quality subjectively. We demonstrate the results in Fig. \ref{fig:r2}. In addition to the binaural dataset (FAIR-Play), we also verify our method on a stereo dataset (YouTube-ASMR). We evaluate our model on the music scenes, the scene with a human whispering, and the scene of a human interacting with objects. It can be observed that the spatial information of visuals and audio is consistent. When the sound source is on the left side of the visual scene, the sound from the left channel is clear.

We compare our results with other network architectures in the ablation studies described in Section 4.3.2, and the results are shown in Fig. \ref{fig:r1}.  We compare our results with the ground-truth and four other architectures (APNet, the backbone network, without attention network, without weighted loss, and attention network). 
We also compare our generated results with the backbone network and APNet on the Youtube-ASMR dataset, as shown in Fig. \ref{fig:asmr}. This is because our backbone network has the same architecture as \cite{gao20192}. To ensure the same experiment settings, we only compare with the backbone network. For \cite{zhou2020sep}, their network is the same as APNet during stereophonic learning, except they further introduce separative learning. As a result, we do not introduce mono audio data in our experiments; instead, we compare our results with those generated by APNet. It can be observed that the left and right audio channels generated by APNet are sometimes similar, and the spatial information of the audio cannot be reconstructed clearly. In contrast, our results do not suffer from these issues and are closer to the ground-truth. It can be found that our generated result is the most similar to the ground-truth. The numeric comparisons for these methods are given in Table 3, which shows the results generated by our method have the smallest distance from the ground-truth.

We also visualize the areas on which the network focuses in the input visual frames. The results are shown in Fig. \ref{fig:vis}. We use visual features $f_{vb}$ obtained in the binaural audio generation task to implement the visualization. If we want to recover the spatial information in the audio, we need to focus more on the areas that contain the sound sources. Fig. \ref{fig:vis} also reflects the same result as our intuition. Values of the envelope distance between the generated results and ground-truth are annotated in Fig. \ref{fig:r1}, which also shows our results have the smallest distance to the ground-truth.

\subsubsection{User study}
 We also conducted a user study to further evaluate our generated results by asking the participants to judge the perceptual quality of our generated sound.
 
 \textbf{Participants.} A total of $20$ participants were involved in our experiments and were generally aged around 20 years. All of them had normal hearing. All participants wore headphones during this procedure.
 
 \textbf{Design \& Procedure.} The participants were shown 5 groups of unmarked videos chosen from the FAIR-Play dataset with audio generated by our method, APNet, and the ground-truth audio. They watched these videos and were asked the following questions:
 \begin{itemize}
 	\item Q1: What is the score assigned to each video?
 	\item Q2: Which video do you prefer?
 \end{itemize}
 The score for each video clip is on a scale from 1 to 5, where 1 corresponds to ``Very bad'' and 5 is ``Very good.'' We asked the participants to evaluate the video mainly from two aspects: the quality of sound and the similarity to the ground-truth. The quality of sound was evaluated mainly from the correspondence of spatial information between video and audio. This measure evaluates the consistency between visual frames and corresponding audio. We also asked the participants to judge the similarity between our results and the ground-truth. From these two aspects, participants scored these videos and chose the better one.
 
\begin{figure}
	\centering 
	\includegraphics[width=\columnwidth]{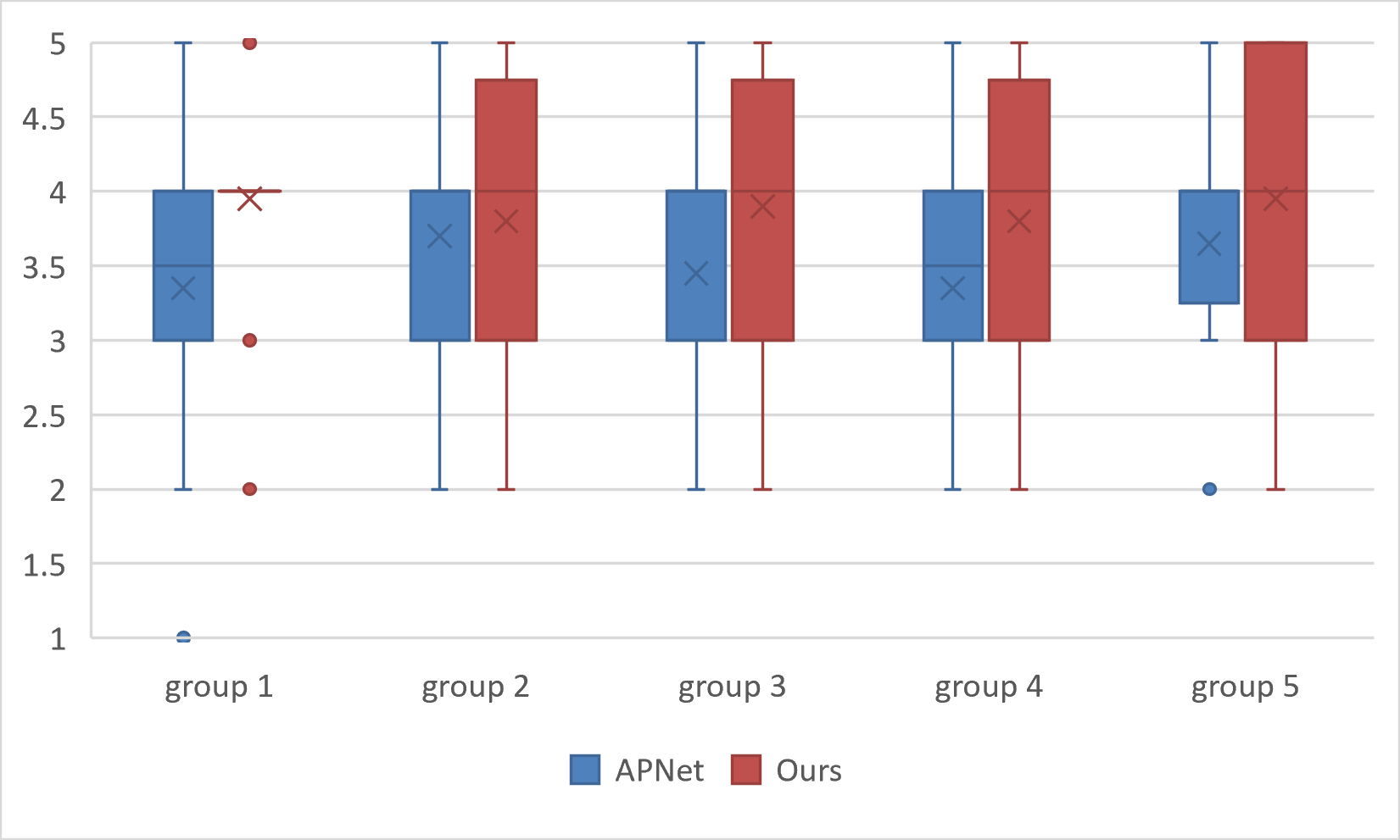}
	\caption{The scores corresponding to the perceptual audio quality generated using APNet and our approach for each group (corresponding to the 5 groups of unmarked videos shown to the participants). Most participants tend to give the same score (4) for our results, causing the box in group 1 to become a line.}
	\label{fig:usr1}
\end{figure}

 \textbf{Results \& Analysis.} We collected the scores given by the participants, and the results are shown in Fig. \ref{fig:usr1}. 
 We also calculated the mean and variance values and show them in Table \ref{tab:usr1}. We observe that our method based on multi-task learning has the highest scores, which shows our results have better quality and are closer to the ground-truth. We also performed the ANOVA test, and results are shown in Table \ref{tab:usr2}. These results indicate the effectiveness of our method. We also asked the participants to choose in terms of better audio results from those generated by our method and APNet. If it was really difficult to choose the better one, they did not report any result. We calculated the percentage that each method was chosen in terms of better audio. Our method accounts for 63\%, APNet for 27\%, and 10\% have the same quality.
\begin{table} 
    \caption{Mean and variance values for the user study.}
    \centering
    \begin{tabular}{p{4cm}<{\centering} p{1cm}<{\centering} p{1cm}<{\centering}}
        \hline
        \hline
        \noalign{\smallskip}
        Method & Mean & Variance  \\ 
        \noalign{\smallskip}
        \hline
        \noalign{\smallskip}
        APNet & 3.5 &  0.8933 \\
        \noalign{\smallskip}
        Ours       &  3.88 & 0.9132  \\
        \noalign{\smallskip}

        \hline
        \hline
        \label{tab:usr1}
    \end{tabular}
\end{table}
\begin{table} 
	\caption{Analysis of the variance. Note that ``SS'' represents sum of square, ``DF'' is degrees of freedom, ``MS'' means mean squares. The last two columns represent F ratio and P values, respectively. We calculate the variance between groups and within groups.}
	\centering
	\begin{tabular}{p{1cm}<{\centering} p{1cm}<{\centering} p{1cm}<{\centering}p{1cm}<{\centering} p{1cm}<{\centering}p{1cm}<{\centering} p{1cm}<{\centering}p{0.5cm}<{\centering} p{0.5cm}<{\centering}}
		\hline
		\hline
		\noalign{\smallskip}
		Groups & SS & df & MS &F &prob>F \\ 
		\noalign{\smallskip}
		\hline
		\noalign{\smallskip}
		Between & 7.22 &  1 & 7.22 &8.85&0.0033 \\
		\noalign{\smallskip}
		Within      &  161.56 & 198&0.8160 & -&- \\
		\noalign{\smallskip}

		\hline
		\hline
		\label{tab:usr2}
	\end{tabular}
\end{table}

\section{Conclusions, Limitations, and Future Work}

We present a method to generate a binaural audio recording from a mono audio recording and the corresponding visual frames through multi-task learning. Our approach unifies two tasks, i.e., generating binaural audio and distinguishing flipped audio based on visual frames. We share the visual network between the two tasks and use this network to extract appropriate features related to the specific task. We highlight the benefits of training for both tasks at the same time. We perform quantitative and qualitative evaluation and highlight the benefits over prior methods. 

\begin{figure}
	\centering 
	\includegraphics[width=2.5 in]{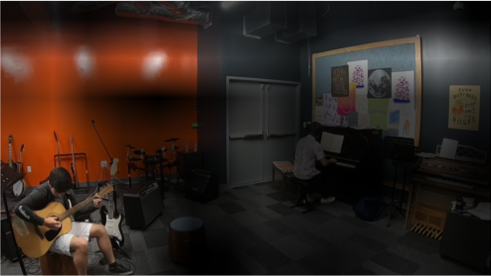}
	\caption{A failure case. There are two sound sources in the scene. However, the bright areas cover only one sound source.}
	\label{fig:fail}
\end{figure}

Our approach has some limitations. First, we need to carefully choose the weights corresponding to different losses manually. Different weights may lead to different results. Therefore, it is important to design good methods to automatically choose the weight. Moreover, the accuracy and perceptual quality of our binaural audio decrease in scenes with multiple sound sources.  A failure case has been shown in Fig. \ref{fig:fail}. There are two sound sources in the scene, and we visualize the salient visual areas learned by our model. The bright areas indicate the more focused regions in the input frame of our network. We can observe from Fig. \ref{fig:fail} that the model focuses more on the sound source on the left and ignores the sound source on the right. Therefore, we need better techniques to handle such complex scenes with multiple sound sources.

There are many avenues for future work. In addition to overcoming the limitations, we would like to evaluate the performance on other recorded videos. We would also like to incorporate temporal information into the framework and develop unsupervised learning methods. In addition, studying how the room properties influence the results and how to improve the generalization are also interesting topics for the future.


\begin{acks}
%
The authors would like to thank the anonymous reviewers for their insightful comments.
Shiguang Liu was supported by the \grantsponsor{GS501100001809}{Natural Science Foundation of China}{} under Grant
No.~\grantnum{GS501100001809}{62072328}
and~\grantnum{GS501100001809}{61672375}.
\end{acks}

\bibliographystyle{ACM-Reference-Format}
\bibliography{Bibliography}


\end{document}